\documentstyle[12pt]{article}
\newcommand{\eqreset}{\setcounter{equation}{0}}
\newcommand{\bgeq}{\begin{equation}}
\newcommand{\eneq}{\end{equation}}
\newcommand{\bgeqa}{\begin{eqnarray}}
\newcommand{\eneqa}{\end{eqnarray}}
\newcommand{\bgit}{\begin{itemize}}
\newcommand{\enit}{\end{itemize}}
\newcommand{\bgen}{\begin{enumerate}}
\newcommand{\enen}{\end{enumerate}}

\newcommand{\braket}[1]{\langle #1 \rangle}

\begin{document}
\title{{\bf
Ground-State Phase Diagrams\\
of the Two-Dimensional Quantum\\
Heisenberg Spin Glass Models\footnote{J.\ Phys.\ Soc.\ Japan\
{\bf 64} (1995) Vol.\ 8.}
}}
\author{
{\Large Yoshihiko~N{\sc onomura} and Yukiyasu~O{\sc zeki}}\\
{\it Department of Physics, Tokyo Institute of Technology,}\\
{\it Oh-okayama 2-12-1, Meguro-ku, Tokyo 152, Japan}
}
\date{}
\maketitle
\noindent
\begin{abstract}
Ground-state properties of the two-dimensional $S=1/2$ random Heisenberg
models are investigated by the exact-diagonalization method. The phase
diagram of the bond-random model (the $\pm J$ model) is the same as that of
the corresponding Ising spin glass model, in spite of quantum fluctuation.
In the site-random model, the spin glass phase exists at least in the
ferro-rich case. In comparison with the corresponding Ising model, the
N\'eel phase of this model becomes much narrower in the ferro-rich case,
while it is comparable in the antiferro-rich case.
\end{abstract}
\eqreset
While various aspects of quantum effects in random spin systems have been
investigated, the recent progress in the random {\it quantum} Heisenberg
model, the most realistic model for spin glass materials, is restricted
to studies in one dimension.\cite{dsf1,dsf2,ngft} Quite recently, one of
the present authors (Y.~N.) investigated\cite{qmat} the $S=1/2$
asymmetric Heisenberg Mattis model in two dimensions.\cite{mattis,asmat}
He found that its ground-state phase diagram does not change\cite{qmat}
from that of the corresponding Ising Mattis model,\cite{asmat} in spite
of quantum fluctuation. He also found that accurate calculations are
possible only using quite small clusters (up to the $20$-spin one)
in the antiferromagnetic region.

In classical spin glass systems, unexpected behavior was reported for
the {\it site-random} Ising spin glass model in two dimensions\cite{sris}
by Shirakura {\it et al.}\cite{2dsris} They analyzed this model using
Monte Carlo simulations, and found that the spin glass phase may be stable
at finite temperatures. The present authors investigated\cite{sristr} the
ground state of the same model in detail using the numerical transfer-matrix
method,\cite{numtr1} and found that the ferromagnetic and the N\'eel phase
regions are wider than the expected values obtained from the corresponding
$\pm J$ model. They also proposed that such structure of the phase diagram
can be understood well in terms of the site-percolation picture.

On the basis of these findings, we investigate ground-state phase
diagrams of several two-dimensional $S=1/2$ random Heisenberg
models described by the following Hamiltonian,
\bgeq
  {\cal H}=-\sum_{\braket{ij}}J_{ij}\vec{S}_{i}\cdot\vec{S}_{j}\ .
\eneq
Properties of the models depend on the choice of randomness in $J_{ij}$. We
consider the conventional bond-random model (the $\pm J$ model)\cite{pmJrev}
in which the distribution function of $J_{ij}$ is given by
\bgeq
  P(J_{ij};p)=p\delta(J_{ij}-1)+(1-p)\delta(J_{ij}+1),
\eneq
and the site-random model\cite{2dsris,sris,sristr} defined through
\bgeq
  J_{ij}=\frac{J}{2}\left[\epsilon(1-\omega_{i}\omega_{j})
                          +\omega_{i}+\omega_{j}\right]\ ,
\eneq
where $\omega_{i}$ is an independent random variable with the distribution
\bgeq
  P(\omega_{i};c)=c\delta(\omega_{i}-1)+(1-c)\delta(\omega_{i}+1)\ .
\eneq
The latter model is characterized by the Ising variable $\omega_{i}$
located at each site. This parametrization represents mixed crystals;
i.e.\ $\omega_{i}=+1$ and $\omega_{i}=-1$ correspond to different ions.
In the case $\epsilon=+1$, $J_{ij}=+J$ unless $\omega_{i}=-1$ and
$\omega_{j}=-1$. We call this case ``ferro-rich" (F-rich). Similarly,
in the ``antiferro-rich" (AF-rich) case ($\epsilon=-1$), $J_{ij}=-J$
unless $\omega_{i}=+1$ and $\omega_{j}=+1$. Note that these two cases
are not equivalent in quantum spin systems.

We have two purposes in the present study. One is to clarify effects
of quantum fluctuation in spin glass systems. Since quantum fluctuation
plays an important role in the antiferromagnetic region, we calculate
the N\'eel order parameter $m_{{\rm st}}$ defined in
\bgeq
  \label{neelop}
  m_{{\rm st}}^{2}\equiv\frac{1}{N^{2}}\sum_{i,j}(-1)^{|i-j|}
    \left[\braket{\vec{S}_{i}\cdot\vec{S}_{j}}\right]_{{\rm r}}\ .
\eneq
Our other purpose is to investigate the difference between bond-random
systems and site-random systems. As in the Ising spin glasses, we consider
the spin-glass (SG) order parameter $m_{{\rm sg}}$ defined in
\bgeq
  \label{sgop}
  m_{{\rm sg}}^{2}\equiv\frac{1}{N^{2}}\sum_{i,j}
    \left[\braket{\vec{S}_{i}\cdot\vec{S}_{j}}^{2}\right]_{{\rm r}}\ .
\eneq

In Ising systems, the lower critical dimension $d_{{\rm lc}}$ of the SG
phase has been expected to be in the range $2<d_{{\rm lc}}<3$\cite{pmJlcd}
in the bond-random model, while it is expected to be equal to that of the
pure Ising model, $d_{{\rm lc}}=1$,\footnote{If the existence of the SG
order parameter at finite temperatures is established in two dimensions
as pointed out by Shirakura {\it et al.},\cite{2dsris} $d_{{\rm lc}}=1$
is trivial. The one-dimensional nearest-neighbor spin glass model is
not frustrated, and can be identified with the ferromagnetic Ising
model by means of a gauge transformation.} in the site-random model.
In quantum Heisenberg systems, though $d_{{\rm lc}}$ has been reported
to be greater than three\cite{3dheisim,3dxysg} in the bond-random model,
the above argument suggests that it is equal to that in the pure quantum
Heisenberg model, $d_{{\rm lc}}=2$, in the site-random model.
If this is the case, the SG order of the three-dimensional
site-random Heisenberg model is expected to be stable even at
finite temperatures, which is consistent with experimental results.

We use the exact-diagonalization method in the present letter. Although
the size of treatable clusters is quite small in this method, plausible
estimation is expected in the antiferromagnetic region, as in the case
of the asymmetric quantum Heisenberg Mattis model.\cite{qmat}
Calculations are carried out on the $8$,$10$,$16$,$18$,$20$-spin
Oitmaa-Betts-type\cite{ob} clusters. In the site-random model, we average
all the configurations. In the bond-random model, we average all the bond
configurations in the $8$ and $10$-spin clusters, and average one thousand
randomly chosen samples in the $16$, $18$ and $20$-spin clusters.

First, we show the data of the N\'eel order parameters. These quantities
in the bond-random, the AF-rich site-random and the F-rich site-random models
are plotted in Figs.\ \ref{neelfig}(a), \ref{neelfig}(b) and \ref{neelfig}(c),
respectively. They are scaled well by $N^{-1/2}$, as in the two-dimensional
pure antiferromagnetic Heisenberg model ($p=0$ or $c=0$ case).\cite{swcorr}
By the least-squares fitting of the $10$,$16$,$18$,$20$-spin data,
we estimate the critical concentrations of the N\'eel phase as
\bgeqa
  \label{brsgcon}
  p_{{\rm c}}&=&0.11\pm 0.01\ \ \ \mbox{(bond-random model)}\ ,\\
  \label{arsrcon}
  c_{{\rm c}}&=&0.36\pm 0.03\ \ \ \mbox{(AF-rich site-random model)}\ ,\\
  c_{{\rm c}}&=&0.13\pm 0.02\ \ \ \mbox{(F-rich site-random model)}\ .
  \label{frsrcon}
\eneqa
In the corresponding Ising spin glass models,
these critical concentrations were estimated as
$p_{{\rm c}}=0.11\pm 0.01$,\cite{numtr1,2dpmJ} $c_{{\rm c}}=0.37\pm 0.01$
and $c_{{\rm c}}=0.41\pm 0.01$,\cite{sristr} respectively.

In the bond-random Ising model ($\pm J$ Ising model), the N\'eel phase
boundary in the $p$-$T$ plane is vertical to the $p$-axis.\cite{2dpmJ,sgver}
Namely, the N\'eel order is not destroyed by thermal fluctuation in the
low-temperature region. If the effect of quantum fluctuation is similar
to that of thermal fluctuation, the equivalence of $p_{{\rm c}}$ in the
Heisenberg model and the Ising model can be understood. Numerically, the
estimates of $c_{{\rm c}}$ are also consistent with each other in the
AF-rich site-random Heisenberg model and Ising model. Although verticality
of the phase boundary has been shown only in the bond-random model, a similar
mechanism might work for the N\'eel order of the AF-rich site-random model.
On the other hand, the estimates of $c_{{\rm c}}$ are quite different in the
F-rich site-random model ($c_{{\rm c}}\sim 0.13$ in the Heisenberg model,
while $c_{{\rm c}}\sim 0.41$ in the Ising model). This nontrivial quantum
effect cannot be explained by the above thermal-quantum equivalence.
A study of the origin of such large difference is now in progress.

\begin{figure}[h]
\caption{
The estimates of the SG order parameter in the thermodynamic limit in
the bond-random model (solid line with crosses), the AF-rich site-random
model (solid line with circles), the F-rich site-random model (solid line
with diamonds) and the asymmetric Mattis model (broken line).
}
\label{sgvalfig}
\end{figure}
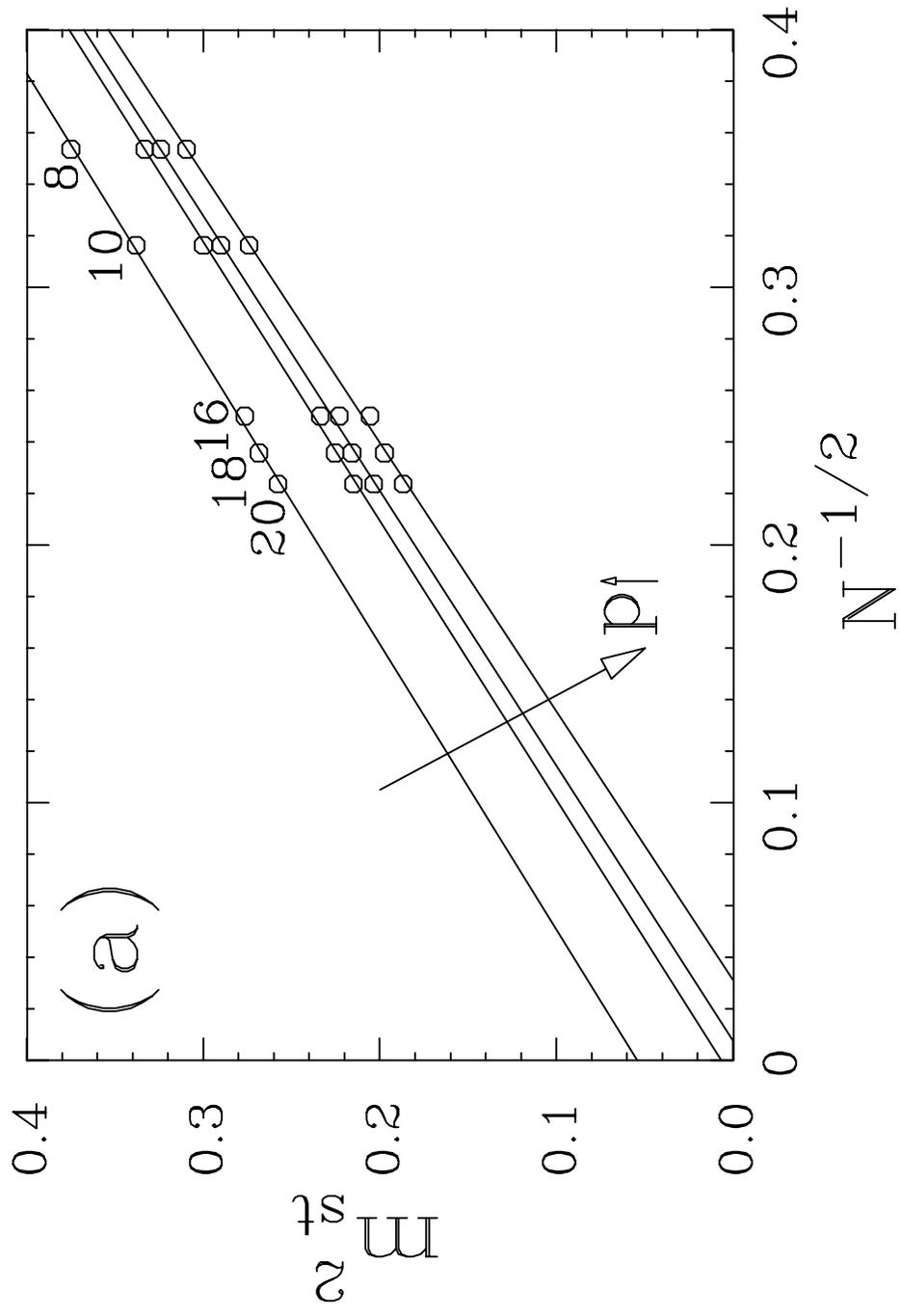
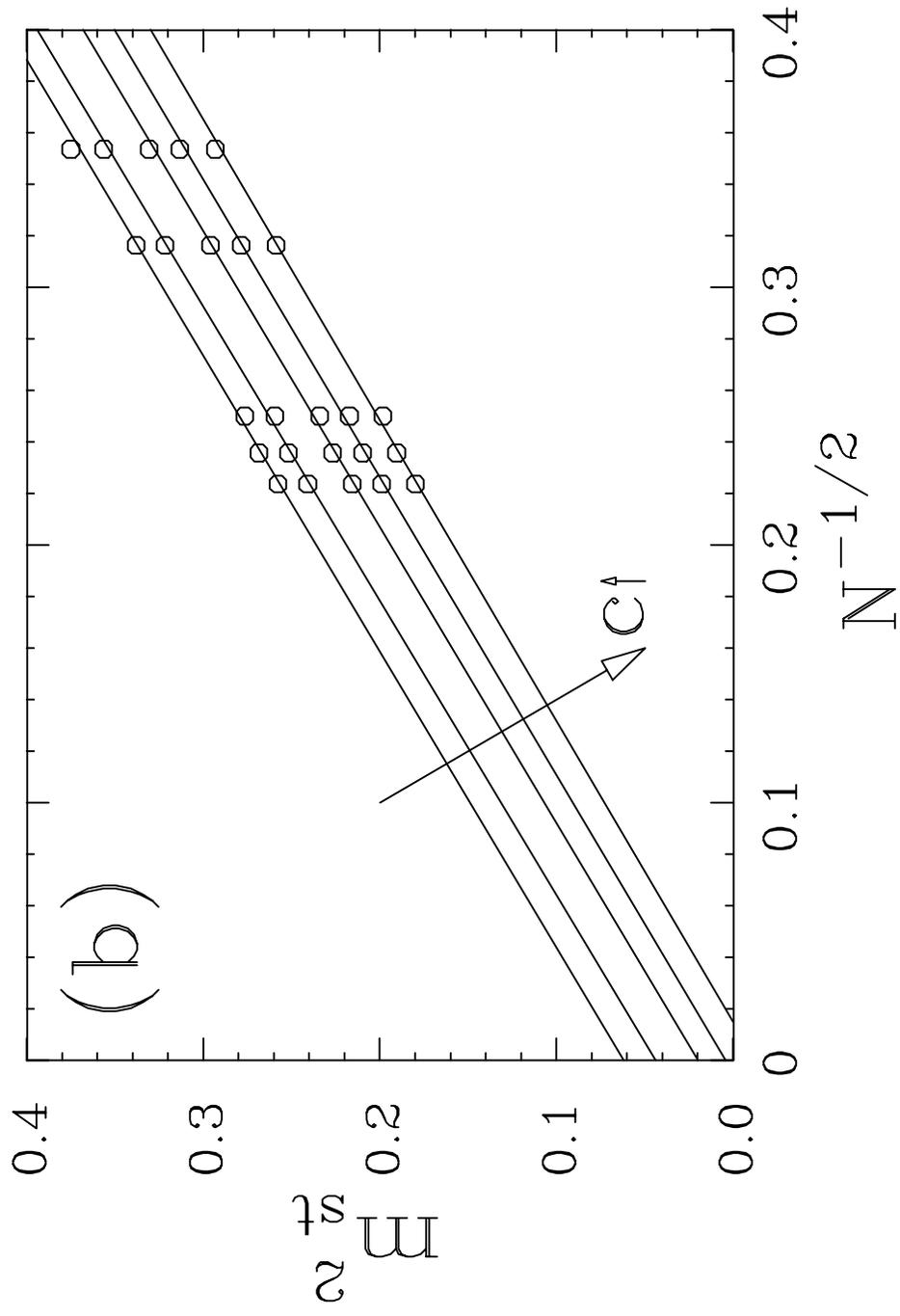
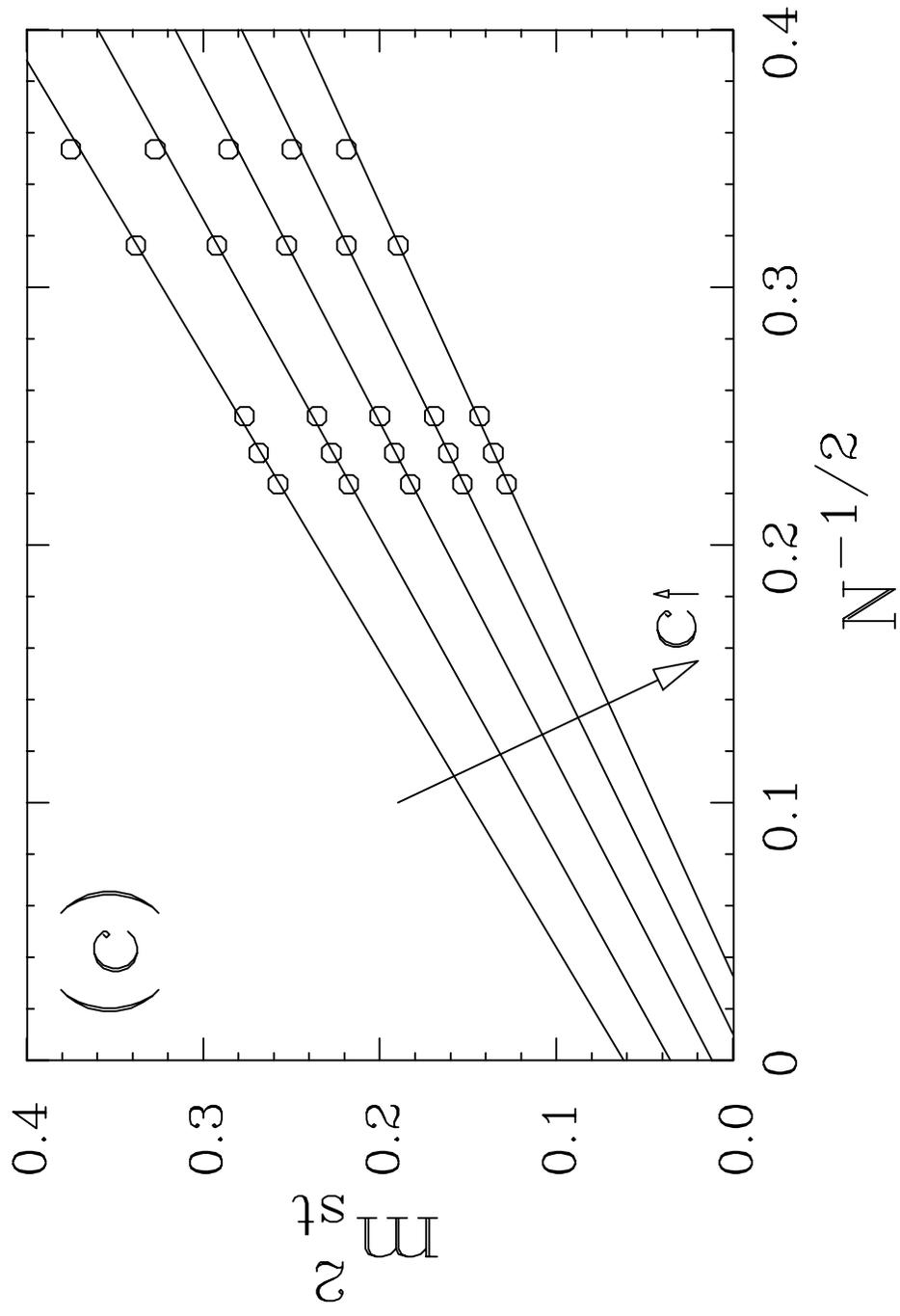
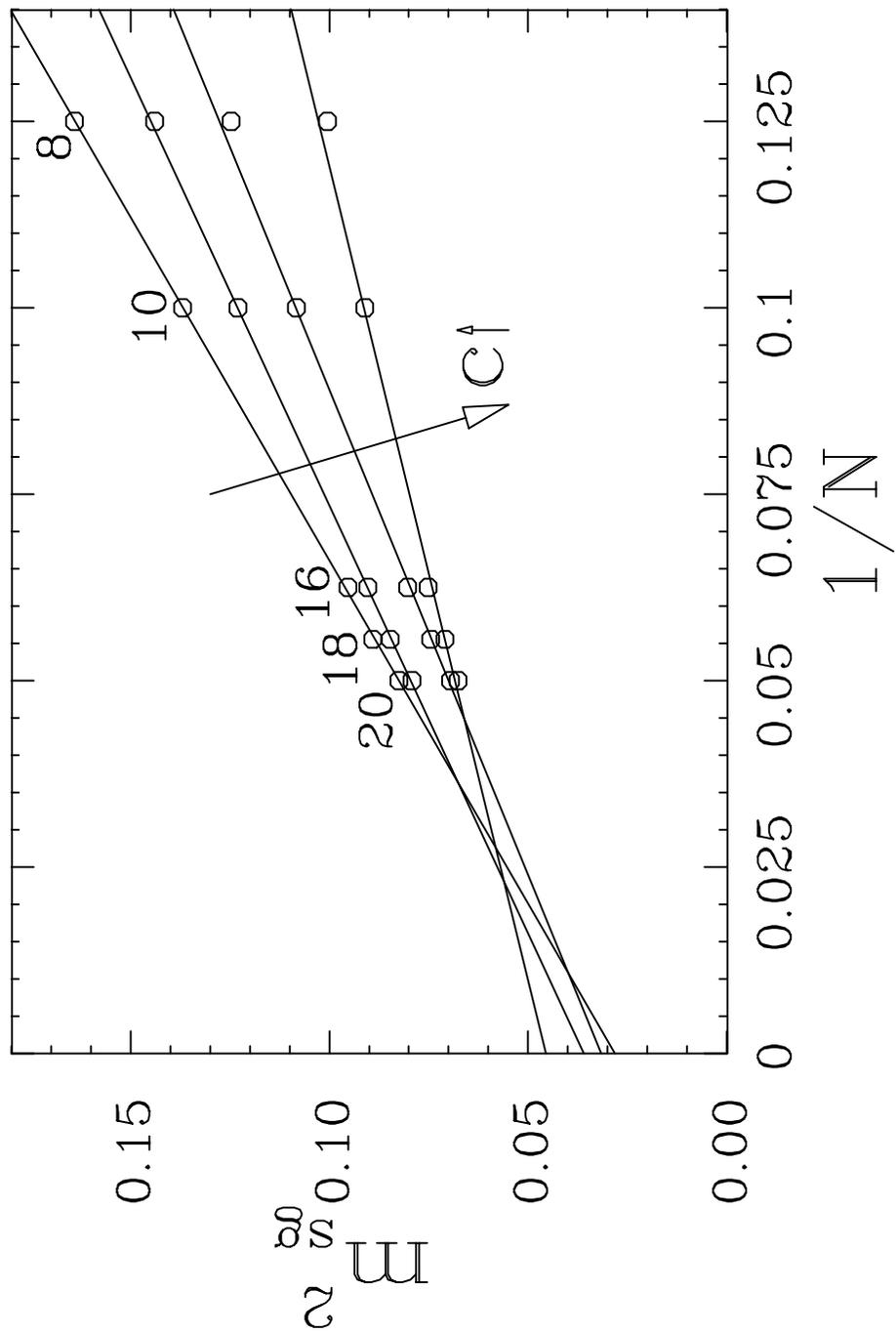
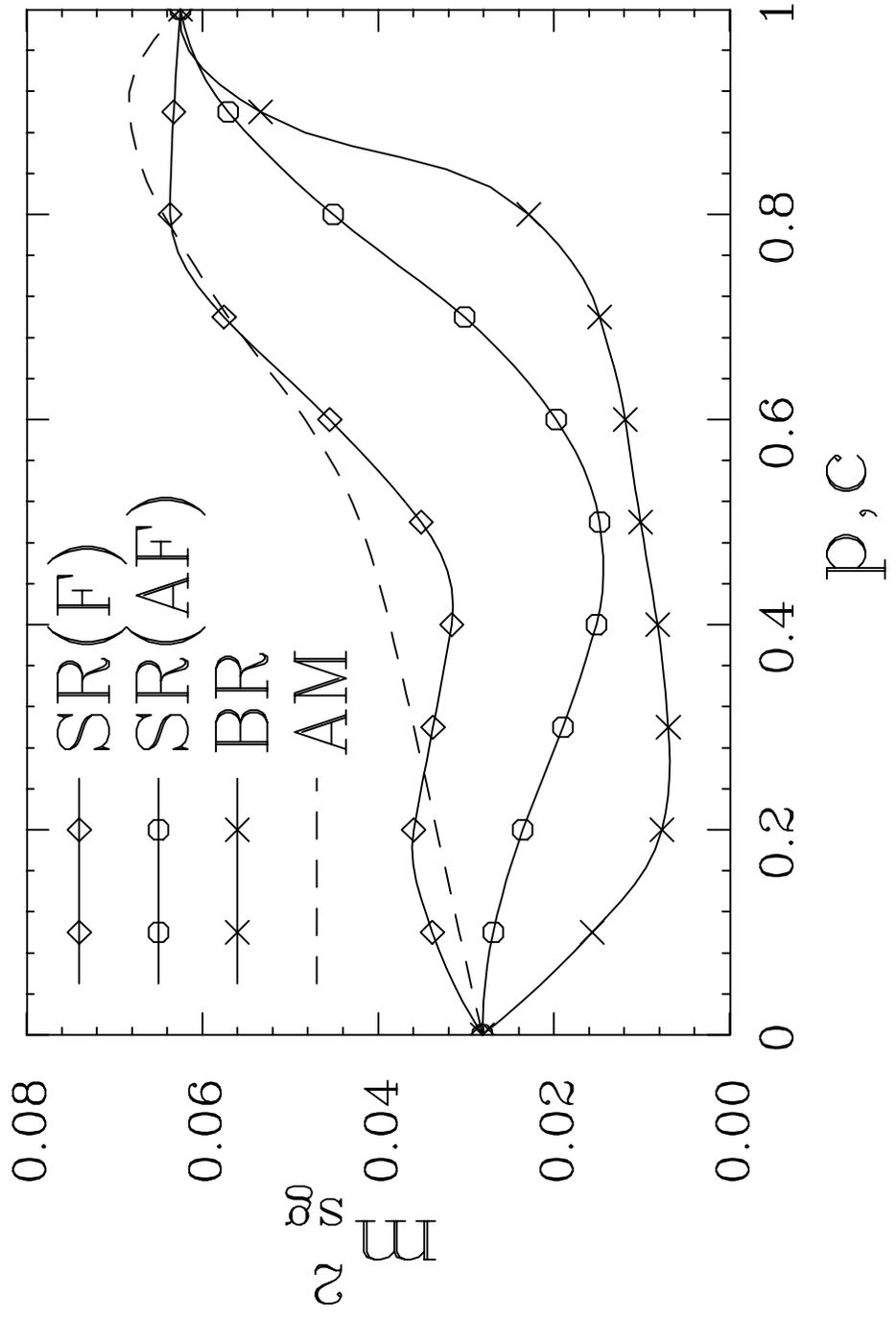

Next, we show the data of the SG order parameters. These quantities
are scaled by $1/N$, as pointed out in the case of the two-dimensional
Heisenberg Mattis model.\cite{qmat} An example of the fitting is given in
Fig.\ \ref{sgfig} for the F-rich site-random model. Similar good scaling
behavior is also observed in the other models. The estimates of the SG
order parameter are plotted versus $p$ or $c$ in Fig.\ \ref{sgvalfig}.
These results suggest that the SG order parameter in the site-random model
is nonvanishing in the whole parameter region of the concentration $c$
in both the F-rich model and the AF-rich model . However, even in the
bond-random model, the SG order parameter also seems to be nonvanishing.
In the two-dimensional bond-random Ising model, this order parameter
is stable only in the vicinity of $p=0$ and $p=1$\cite{numtr1,2dpmJ}
in the ground state. This should also be the case in the present
bond-random Heisenberg model, in which the SG order is weakened
more by quantum fluctuation. Thus, there still remains a finite-size
effect in the present calculation, and the existence of the SG order
cannot be concluded only from these results. In fact, in the AF-rich
site-random model, the estimate of the SG order parameter decreases
as $c$ increases, and becomes as small as those in the bond-random
model around $c\sim 0.5$.

The estimate of the SG order parameter in the Heisenberg Mattis
model\cite{qmat} is also displayed in Fig.\ \ref{sgvalfig}. The previous
study of this model\cite{qmat} showed that the SG order parameter is
nonvanishing at all concentrations. The concentration dependence in
the F-rich site-random model is similar to that in the Mattis model,
except for small reduction around $c\sim 0.5$. Even in this region,
the estimate of the order parameter is larger than that at $c=0$.
Thus, we can at least claim that the SG order remains nonvanishing
in the F-rich site-random Heisenberg model in two dimensions.

In summary, we study ground-state phase diagrams of various
$S=1/2$ Heisenberg spin glass models in two dimensions. In the
bond-random model (the $\pm J$ model), the critical concentration
of the N\'eel phase is estimated as $p_{{\rm c}}=0.11\pm 0.01$,
which is consistent with the corresponding Ising spin glass model.
Such equivalence is expected to be related to the vertical phase
boundary in the phase diagram of the bond-random Ising model.
In the site-random model, the critical concentration is consistent with
the Ising spin glass model in the AF-rich case ($c_{{\rm c}}=0.36\pm 0.03$),
while that in the F-rich case ($c_{{\rm c}}=0.13\pm 0.02$) is decreased much
more than that of the corresponding Ising model ($c_{{\rm c}}\sim 0.41$).
This reduction is the only clear quantum effect displayed in the present
study. Moreover, the evidence for the existence of the spin glass phase
is found in the two-dimensional quantum Heisenberg site-random model in
the ground state, at least in the F-rich case. Studies on the mechanism of
the large reduction of the N\'eel phase in the F-rich site-random model and
structures of excitation spectra in the present models are now in progress.

One of the present authors (Y.\ N.) is grateful for the financial support
of the Japan Society for the Promotion of Science for Japanese Junior
Scientists. This work was supported by Grant-in-Aid for Encouragement
of Young Scientists, from the Ministry of Education, Science and Culture.
The computer programs are based on the Hamiltonian-diagonalization
subroutine package ``TITPACK Ver.\ 2" developed by Professor H.~Nishimori.
The numerical calculations were performed on HITAC S3800/480
at the Computer Center, University of Tokyo and on FACOM VPP 500
at the Institute of Solid State Physics, University of Tokyo.
\noindent
\section*{Figure Captions}
\noindent
\begin{figure}[h]
\caption{
The N\'eel order parameters of $8$,$10$,$16$,$18$,$20$-spin clusters
are plotted versus $N^{-1/2}$ (a) in the bond-random Heisenberg model
for $p=0.0,0.1,0.12,0.15$ (from above to below), (b) in the AF-rich
site-random Heisenberg model for $c=0.0,0.2,0.3,0.35,0.4$ and (c) in
the F-rich site-random Heisenberg model for $c=0.0,0.05,0.1,0.15,0.2$.
The straight lines are drawn by the least-squares fitting of the last
four data of each set.
}
\label{neelfig}
\end{figure}
\noindent
\begin{figure}[h]
\caption{
The SG order parameters of $8$,$10$,$16$,$18$,$20$-spin clusters are
plotted versus $1/N$ in the F-rich site-random Heisenberg model for
$c=0.0,0.2,0.4,0.6$ (from above to below). The straight lines are
drawn by the least-squares fitting of the last four data.
}
\label{sgfig}
\end{figure}
\noindent
\end{document}